\documentclass{article}

\usepackage{arxiv}

\usepackage[utf8]{inputenc} 
\usepackage[T1]{fontenc}    
\usepackage{hyperref}       
\usepackage{url}            
\usepackage{booktabs}       
\usepackage{amsfonts}       
\usepackage{nicefrac}       
\usepackage{microtype}      
\usepackage{lipsum}		
\usepackage{graphicx}

\usepackage{array}
\usepackage{amssymb, amsmath, amsthm}
\usepackage{graphicx}
\usepackage{lmodern,url}
\usepackage{makecell} 
\usepackage{cancel}
\usepackage{multirow}
\usepackage{microtype}
\usepackage{lineno}
\usepackage{xspace}
\usepackage{xcolor}
\usepackage{siunitx}
\usepackage{todonotes}
\usepackage{booktabs}

\usepackage{mathdots}
\usepackage{subcaption}
\newcommand{\dis}{\displaystyle}

\newcommand{\bs}[1]{\boldsymbol{#1}}

\title{A multi-group SEIRA model for the spread of COVID-19 among heterogeneous populations}

\usepackage{authblk}

\author[1,2]{Sebastián Contreras}
\author[2]{H. Andr\'es Villavicencio}
\author[2,3]{David Medina-Ortiz}
\author[2,4]{Juan Pablo Biron-Lattes}
\author[2,4\thanks{\tt{aolivera@ing.uchile.cl}}]{\'Alvaro Olivera-Nappa}
\affil[1]{Laboratory for Rheology and Fluid Dynamics, Universidad de Chile, Beauchef 850, 8370448 Santiago, Chile.}
\affil[2]{Centre for Biotechnology and Bioengineering, Universidad de Chile, Beauchef 851, 8370448 Santiago, Chile.}
\affil[3]{Division of Chemistry and Chemical Engineering, California Institute of Technology, Pasadena, CA 91125, USA}
\affil[4]{Department of Chemical Engineering, Biotechnology and Materials, Universidad de Chile, Beauchef 851, 8370448 Santiago, Chile.}

\date{}

\renewcommand{\headeright}{ }
\renewcommand{\undertitle}{ }

\begin{document}
\maketitle

\begin{abstract}
The outbreak and propagation of COVID-19 have posed a considerable challenge to modern society. In particular, the different restrictive actions taken by governments to prevent the spread of the virus have changed the way humans interact and conceive interaction. Due to geographical, behavioral, or economic factors, different sub-groups among a population are more (or less) likely to interact, and thus to spread/acquire the virus. In this work, we present a general multi-group SEIRA model for representing the spread of COVID-19 among a heterogeneous population and test it in a numerical case of study. By highlighting its applicability and the ease with which its general formulation can be adapted to particular studies, we expect our model to lead us to a better understanding of the evolution of this pandemic and to better public-health policies to control it.
\end{abstract}


\section{Introduction}

Since the outbreak of novel SARS-CoV2 in late 2019, the world's reaction capacity has been challenged in every possible way. From the first self-isolation measures to total lock-downs and quarantines for preventing its spreading, the way humans interact has changed, and probably, for good. Joined efforts from the scientific community have helped to design strategies for supporting public-health policies, and for providing a better understanding of the current scenario. In particular, several mathematical models have been proposed in recent weeks to fit public databases on the SARS-CoV2 outbreak and forecast its evolution. Despite their particularities, most of them evolve from the well-known SIR model proposed by \cite{kermack1927contribution}. The basic idea behind this model, and its variants, is to divide a population in different compartments, representing the $S$ (susceptible), $E$ (exposed), $I$ (infected), $A$ (asymptomatic), $R$ (recovered), $D$ (dead), $Q$ (quarantined), among other fractions of it.  Centering our analysis on the spread of COVID-19, and restricting our search to those models recently proposed, we may find different combinations of the variables mentioned above. 

Even though research has been done on SIR and SIRD models for COVID-19, with different levels of complexity--- as in \cite{biswas2020covid,simha2020simple,chen2020time}, among others, for SIR, and in \cite{calafiore2020modified,fanelli2020analysis,bastos2020modeling}, among others, for SIRD---, properties of the virus and the way modern society interacts are more likely to be appropriately represented by SEIR/SEIRA models. 

\cite{yang2020modified} presented a modified SEIR model, where variables $S$ and $E$ had an in-out function, accounting for quarantine, and $E$ was considered to be infectious. Even though the model involved several variables, the parameter-fitting stage only considered an exponential approximation, leaving unclear how the infection and recovery rates were decoupled. Another variant of SEIR models was presented by \cite{yang2020mathematical}, including an equation for the concentration of SARS-CoV2 in the environmental reservoir. \cite{peng2020epidemic} present a sophisticated 7-compartment variant of SEIR models, where the extra variables account for the quarantined $Q$, dead $D$, and insusceptible $P$ fractions of the population. Moreover, their model includes the temporal evolution of the rates of recovery and infection-driven death. \cite{kucharski2020early} used a stochastic dynamic model to fit publicly available datasets of Wuhan to give better insights on the early dynamics of COVID-19, and therefore predict if newly introduced cases would generate outbreaks in other areas. \cite{pang2020public} present several public health recommendations, based on the modeling of the effect of different regulations for preventing the spread of COVID-19 on a SEIRA model similar to the one presented by \cite{aguilar2020investigating}, highlighting the challenges that an epidemic of an aerial contagious virus with asymptomatic patients pose. 

Nevertheless, the different models listed above do not account for the several factors that could result in a heterogeneous spread of the infection in the population. Barriers can be risen by the geo-spatial configuration of a country and differential connectivity between its regions, or even by the way the different communities or social classes interact in a city. Even though work has been done in multi-group SIR models \cite{hethcote1985stability} and the stability of their endemic equilibrium \cite{guo2006global,sun2011global}, the particularities of the COVID-19 pandemics and human interaction networks require to be modeled to be correctly represented.

In this work, we present a general multi-group SEIRA model for representing the spread of novel COVID-19 through populations with heterogeneous characteristics, such as a heavily centralized organization with poor connections between the provinces, substantial social inequality among the population, or age/behavioral groups. Due to the generality of our model, we can easily modify it to account for different particularities of any population. In our model, we represent the interactions between the different groups by a non-necessarily symmetrical matrix connecting the different sub-groups, accounting for the fraction that effectively interacts. Particularities of novel COVID-19 were used to model the infection dynamics, so individuals could be infected by foreigners in their region, or while they are visiting a different one. We apply our model in a case of study for representing the dynamics between different behavioral groups, highlighting its applicability to the early prediction of contingencies.

\section{Our approach}

The foundations on which our model relies are the well-known mathematical and biological hypotheses for SEIR single and multi-group models \cite{hethcote1985stability}. Other suppositions, as listed below, are based on the state-of-the-art knowledge of COVID-19 and other population-balance considerations.

\begin{itemize}
    \item There exists an incubation period for the virus, which average has been reported to range between 5.2 days \cite{li2020early, lauer2020qifang} and 6.4 days \cite{backer2020incubation}. This fact justifies the existence of a compartment $E$ (exposed).
    \item The asymptomatic fraction of a given population, which varies from 29\% \cite{li2020clinical}, going through 50\% \cite{mizumoto2020estimating} and rising between 50 and 75\% \cite{day2020covid}, will be considered constant~$\alpha$. This hypothesis makes sense if the size of the population and its immunological characteristics are homogeneous. 
    \item The only possible interaction among different classes is infection. Recoveries and deaths are supposed to occur in the group where they belong.
    \item Asymptomatic patients are $\xi$ times more likely to transmit the infection successfully to a healthy patient. Even though the viral load of asymptomatic individuals is in the same order of magnitude than in symptomatic patients \cite{zou2020sars}, our supposition relies on a social-interaction basis. Symptomatic patients are easily avoided, quarantined, or likely to choose self-isolation, which is not the case of asymptomatic individuals.
\end{itemize}

Our model also accounts for Asymptomatic individuals $A$, but as we suppose them to represent a constant $\alpha$ fraction of the total infected individuals, they are considered in the equation for $I$. 

\begin{equation}
    I_{\text{total}} = \underbrace{(1-\alpha) I}_{\text{Symptomatic}} + \overbrace{\alpha I}^{A: \text{ Asymptomatic}}.
\end{equation}

\begin{figure*}[ht!]
    \centering
    \includegraphics[scale=.4]{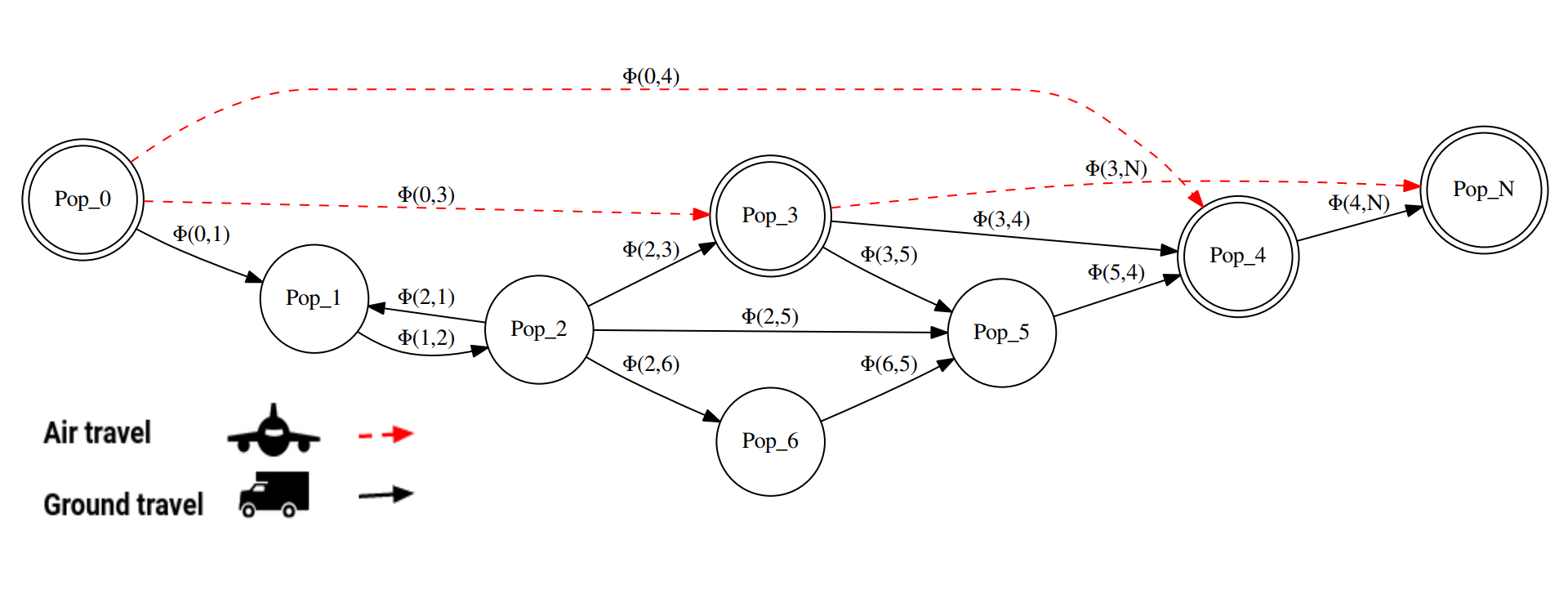}
    \caption{Our approach proposes the partition of a heterogeneous population into several (as-many-as-required) sub-populations, where the hypotheses for SEIR models are satisfied. The different populations share common characteristics, as a geographical zone (this scheme), but not restricted only to that interpretation. This same reasoning (and model) can be applied to different behavioral groups, social classes, and age groups, through an appropriate interpretation for the interaction function.}
    \label{EsquemaInteraccionCajas}
\end{figure*}

The general idea behind our model, and workflow of its network-scheme, is presented in Figure~\ref{EsquemaInteraccionCajas}. The particular deduction and novelties involved in the SEIR equations are presented in the following section.

\section{The model}

As the different sub-groups interact, we expect to find in the $i$'th group representatives of the other groups. As each group is subdivided into classes according to the SEIRA model, the total amount of a given $X$ class, $X=\{S,E,I,R,n\}$, present in a group $i$ is given by~\label{equation}:

\begin{equation}
    X_i^T = \underbrace{X_i\left(1-\sum_{j\neq i}^{n}\Phi_X^{ij}\right)}_{\text{Locals staying in their own district}} + \underbrace{\sum_{j\neq i}^{n}\Phi_X^{ji}X_j}_{\text{visitors (floating population)}},
\end{equation}

where $\Phi^{ji}$ represents the fraction of the $j$'th class present in class $i$. As the maximum amount of individuals that may leave their class is the total of it, the following restriction must be met:

\begin{equation}\label{condicion}
\sum_{j\neq i}^{n}\Phi^{ij} \leq 1.
\end{equation}

By defining an ``auto-fraction'' $\Phi_X^{ii}$, the notation is considerably enlightened, as a single term would account for the whole population.

\begin{equation}\label{autoflujo}
    \Phi_X^{ii} = 1-\sum_{k\neq i}\Phi_X^{ij},\Rightarrow X_i^T = \sum_{j=1}^{n}\Phi_X^{ji}X_j.
\end{equation}

For studying variations of the susceptible class among the different populations, we must consider the factors which modify it:

\begin{equation}
    \frac{dS_i}{dt}  = \text{Natural net-growth} - \text{Intra-class contagion} - \text{Inter-class infections},
\end{equation}

\begin{figure}[ht!]
    \centering
    \includegraphics[scale=.6]{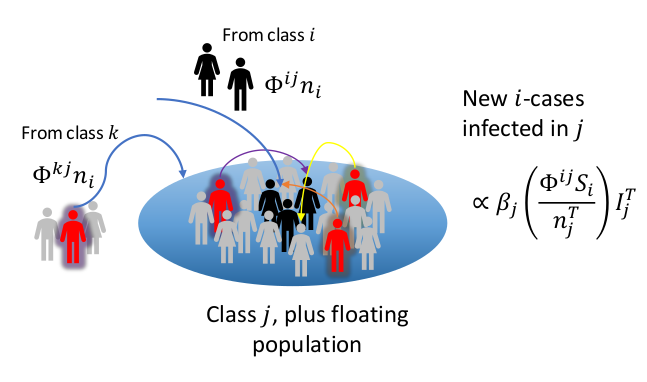}
    \caption{Schematic representation of the contagion processes between classes. According to the form of the interaction matrix $\Phi$, individuals from class $i$ would contribute to the floating population of class $j$, where they would interact with individuals from -in principle- all other classes. As among them might be infected individuals from all origins, the interaction term should be corrected as presented.}
    \label{Esquemacontagios}
\end{figure}

Only with the aim of providing a general framework, easily extendable to other situations, we can also consider the natural birth-death dynamics of the population, characterized by $\Lambda,\,p_i$, and $d_i$. Note that we can neglect those terms if the dynamics of the infection are faster than them. The interesting part is in the dynamics behind contagion, as schematized in Figure~\ref{Esquemacontagios}.

Consider the fraction $\Phi^{ij}n_i$, representing the part of the $i$'th population that is among $j$. There, this group would interact with the total population present at a time (locals and floating population), among whom infected people could be found (red individuals in Figure~\ref{Esquemacontagios}) from all the different classes (the color of their aura). The contagion process can be modeled by equation~\ref{descContagios}

\begin{equation}\label{descContagios}
    \text{New infections} \propto  \underbrace{\beta_i}_{\text{Local infection rate}} \cdot \overbrace{\left(\frac{\Phi^{ij}S_i}{n_j^T}\right)}^{\text{ Density of $i$ in $j$}} \cdot\underbrace{\left(\sum_{k=1}^{n}\Phi^{kj}I_k\right)}_{\text{Total infected in $j$}}.
\end{equation}

Note that equation~\ref{descContagios} also accounts for the intra-class contagion, when defining $\Phi^{ii}$ in accordance to equation~\ref{autoflujo}. In particular, supposing new infections passed by asymptomatic individuals are $\xi$ times more likely to occur, and all the other variables following a typical SEIR scheme, we may write the total dynamics for the $i$'th group:

\begin{align}
    \frac{dS_i}{dt}  &= (1-p_i)\Lambda_i - d_iS_i \qquad -(1+(\xi-1)\alpha)\sum_{j=1}^n\beta_j\left(\frac{\Phi^{ij}S_i}{n_j^T}\right)\left(\sum_{k=1}^{n}\Phi^{kj}I_k\right) \label{dsfinal}\\
    \frac{dE_i}{dt}  &= (1+(\xi-1)\alpha)\sum_{j=1}^n\beta_j\left(\frac{\Phi^{ij}S_i}{n_j^T}\right)\left(\sum_{k=1}^{n}\Phi^{kj}I_k\right)\qquad-\left(\epsilon_i + d_i\right) E_i\\
    \frac{dI_i}{dt}  &= \epsilon_i E_i -(\gamma_i+\theta_i+d_i)I_i\label{dI}\\
    \frac{dR_i}{dt}  &= \gamma_i I_i-d_i R_i
\end{align}

Note that equation~\ref{dI} is the result of considering the dynamics of both symptomatic $((1-\alpha)I)$ and asymptomatic $(\alpha I)$ individuals together.

An important remark on the nature of the interactive term of equation~\ref{dsfinal}, is that the different involved variables are fully coupled:

\begin{align}
    \sum_{j=1}^n(1+(\xi-1)\alpha)\beta_j\left(\frac{\Phi_S^{ij}S_i}{n_j^T}\right)&\left(\sum_{k=1}^{n}\Phi_I^{kj}I_k\right)\\ & = \sum_{k=1}^n\beta_k f\left(S_i,\bs{X},\bs{\Phi}\right).
\end{align}

In the special case of not considering births and deaths from other causes and if all the different fluxes $\dis\Phi^{ij}$ are zero (for~$j\neq i$), function $f$ would be linear for $I_j$. In any other situation, $f\left(S_i,\bs{X},\bs{\Phi}\right)$ would be a non linear function not only of $S_i$ and $I_k$, but of all the involved variables $X_j$. Nevertheless, if the flux structure of the network $\dis\bs{\Phi}$ remains constant (or it is constant by parts), function $f$ satisfies the hypothesis presented in \cite{sun2011global} for the global stability of the endemic equilibrium.

\section{Case of study: Mutual dependency between heterogeneous groups}

Let us assume the actual case of three different behavioral and social groups in Santiago, Chile, A, B, and C, also listed in decreasing order of income and with a high degree of social inequality. Fractions of C are likely to provide services in zones A and B, and therefore interact with their population. Zone B acts as the center of the city, where people meet for office work. Supposing $e$ is the characteristic fraction of the population working in different zones, the interaction matrix $\bs{\Phi_1}$ in normal circumstances \textbf{(Case 1)} is given by equation~\ref{matphi}. Once the outbreak took place in A, its dynamics were controlled by parameters $(\beta,\gamma,\theta)$, which were also valid for its propagation to B. As life quality and access to public-health is worse in C, we will suppose such parameters to be, respectively, half and double from the values reported for them in A. The contagious rate in C would also be higher (for example, a 10\% higher), as the lack of work-safety policies would press people to work even if they already started to feel sick. Then, the dynamics in C would be characterized by $(2\beta,\frac{\gamma}{2},1.1\theta)$. Aware of this consideration, but unable to quantify its effect, the government studied to isolate A partially and to limit its contact with C, generating a new interaction matrix $\bs{\Phi_2}$ \textbf{(Case 2)}:

\begin{equation}\label{matphi}
    \bs{\Phi_1}  = \begin{pmatrix} 1-\frac{e}{2} & \frac{e}{2} & 0 \\ e & 1-2e & e \\ e & 2e & 1-3e  \end{pmatrix},\qquad \bs{\Phi_2} = \begin{pmatrix} 1-\frac{e}{5} & \frac{e}{5} & 0 \\ \frac{e}{5} & 1-\frac{6e}{5} & e \\ 0 & e & 1-e  \end{pmatrix}
\end{equation}

\begin{figure*}[hb!]
     \centering
     \begin{subfigure}[t]{0.48\textwidth}
         \centering
         \includegraphics[scale=.5]{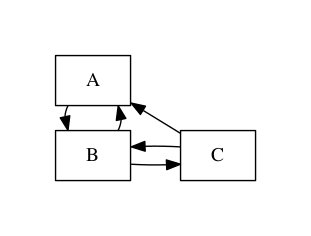}
         \caption{Case 1: Interactions between the different zones follow their regular rate, even though the first cases were already reported.}
         \label{Caso1}
     \end{subfigure}
     \hfill
     \begin{subfigure}[t]{0.48\textwidth}
         \centering
         \includegraphics[scale=.5]{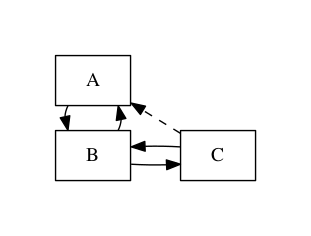}
         \caption{Case 2: Interactions between the different zones were almost suspended. This configuration is also valid for cases 3 and 4.}
         \label{Bar_Plot_N_gp100}
     \end{subfigure}
     \caption{Schematic representation of the two interactive configurations between zones A, B, and C. Zone C has a working-dependency with zone A, and zone B acts as the middle point. This is a simplified version of a case observed in Santiago de Chile, where the outbreak was concentrated in the wealthiest part of the city, but quickly spread to zones where service providers live.}
     \label{Experimento2}
\end{figure*}

Other studied possibilities were the reduction of social interaction by the application of obligatory quarantine after a time $t_c$ from the day of the first infection, generating \textbf{cases 3 and 4}, which considered the evolution of case 1 or 2 respectively. Following the idea of~\cite{del2007mixing}, we assume the rate of contagion is directly proportional to the factor of exposure, so that:

\begin{equation}
    \frac{\beta^q}{\beta} = \frac{f_{\text{exp}}^{1}}{f_{\text{exp}}^{0}}.
\end{equation}

\begin{figure*}[ht!]
    \centering
    \includegraphics[scale=.4]{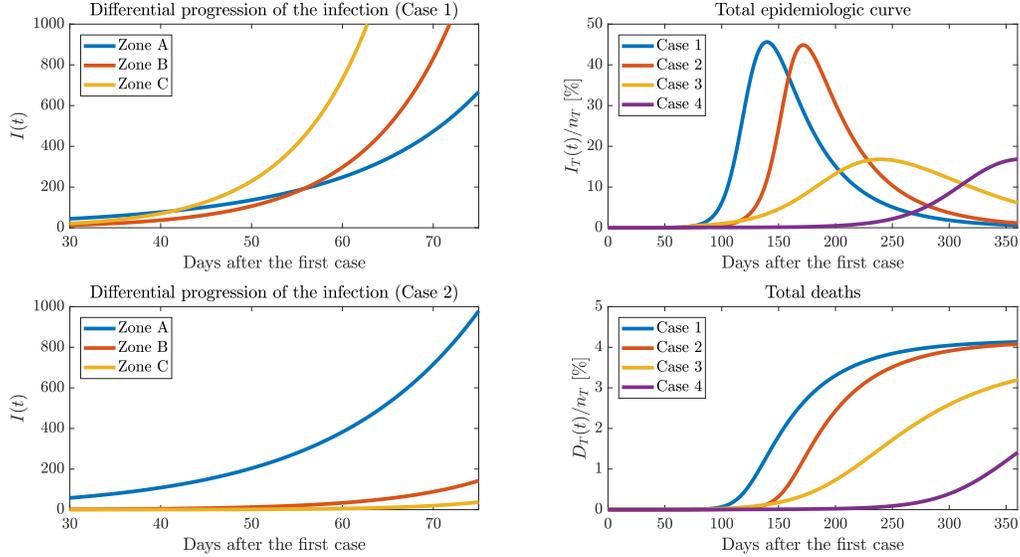}
    \caption{Numeric simulations of cases 1-4, with outbreak parameters $\beta= 0.11$, $\gamma= 0.034$, $\theta= 0.001$, $\xi = 2$, $\alpha= 0.3$, $n = 10^{5}[2.5\, 5\, 6]$, $\beta^q= 0.044$, $I_A(t=0) = 10$. Note that this simulation does not consider the saturation of the local health-system, a parameter that drastically increases the mortality rate of this virus when surpassed.}
    \label{FigNumerica}
\end{figure*}

Figure~\ref{FigNumerica} show the results of numerical simulations of cases 1-4. The comparison of the different scenarios shows that under these considerations our model was able to adequately represent the effect of the dynamics, providing useful insights to lead the decision-making processes when different behavioral groups interact.

\section{Conclusions}

In the present work, we developed a multi-group SEIRA model that is able to represent the spread of COVID-19 in the SARS-CoV2 outbreak through populations with heterogeneous characteristics, which may be given by geographical particularities of the territory, or marked behavioral differences among social classes in a city, country or region. Because of its generality, this model can represent several mechanisms of interaction between different sub-populations and may lead to a better understanding of the evolution of this pandemic, and thus to better public-health policies.

The general structure of the model presented on this work can be easily modified to account for different particularities of given populations or to support policy makers to take public health decisions with greater effectiveness. For instance, the effect of behavioral changes in the population affecting beta, the saturation of the public health system, and the transient dynamics of the fluxes or interactions between sub-populations can be easily implemented in the model by setting the correct functional form in the parameters. As introducing such dynamics would imply adding more parameters, we suggest doing so only if there is enough data to fit them.

Since the way individuals interact is one of the principal aspects of human society that has been affected by this worldwide health emergency, we believe that including interactions in our models and analyses will help us develop strategies that will lead us sooner to recover the richness it used to have.

\section*{Conflict of Interest Statement}

The authors declare that the research was conducted in the absence of any commercial or financial relationships that could be construed as a potential conflict of interest.

\section*{Author Contributions}
Conceptualization, SC, HAV; methodology, SC, HAV; validation AO-N, DM-O, JPB-L; investigation, SC, HAV, DM-O; writing, review and editing, SC, DM-O, HAV, AO-N, JPB-L; supervision, AO-N, SC; project administration, AO-N; funding resources, AO-N.

\section*{Acknowledgements}
The authors gratefully acknowledge support from the Chilean National Agency for Research and development through ANID PIA Grant AFB180004, and the Centre for Biotechnology and Bioengineering - CeBiB (PIA project FB0001, Conicyt, Chile). DM-O gratefully acknowledges Conicyt, Chile, for PhD fellowship 21181435. 

\section*{List of symbols}

\begin{tabular}{cp{0.6\textwidth}}
  $X$ & Arbitrary variable for representing a generic fraction \\
  $n_i$ & Base number of members class $i$ \\
  $n_i^T$ & Effective number of members class $i$ \\
  $\alpha$ & Asymptomatic ratio of the population  \\
  $\xi$ & Extra factor of behavioral virulence of asymptomatic patients  \\
  $\Phi^{ij}$ & Fraction of class $i$ in class $j$ \\
  $p_i$ & Immunity ratio of newborns of class $i$ \\
  $\Lambda_i$ & Net population growth rate $i$\\
  $d_i$ & Per-capita base death rate of class $i$\\
  $\beta_i$ & Infection rate of the virus in class $i$ \\
  $\epsilon_i$ & Inverse of the incubation time in class $i$ \\
  $\gamma_i$ & Recovery rate of class $i$ \\
  $\theta_i$ & Pathogen induced death rate in class $i$\\
  $\bs{\Phi}$ & Interaction matrix \\
  $f_{\text{exp}}$ & Factor of exposure to the infection\\
\end{tabular}

\bibliographystyle{unsrt}


\end{document}